\PassOptionsToPackage{unicode}{hyperref}
\PassOptionsToPackage{hyphens}{url}
\PassOptionsToPackage{dvipsnames,svgnames,x11names}{xcolor}
\documentclass[
  11pt,
  a4paper,
]{article}

\usepackage{amsmath,amssymb}
\usepackage{setspace}
\usepackage{iftex}
\ifPDFTeX
  \usepackage[T1]{fontenc}
  \usepackage[utf8]{inputenc}
  \usepackage{textcomp} 
\else 
  \usepackage{unicode-math}
  \defaultfontfeatures{Scale=MatchLowercase}
  \defaultfontfeatures[\rmfamily]{Ligatures=TeX,Scale=1}
\fi
\usepackage{lmodern}
\ifPDFTeX\else  
\fi
\IfFileExists{upquote.sty}{\usepackage{upquote}}{}
\IfFileExists{microtype.sty}{
  \usepackage[]{microtype}
  \UseMicrotypeSet[protrusion]{basicmath} 
}{}
\makeatletter
\@ifundefined{KOMAClassName}{
  \IfFileExists{parskip.sty}{%
    \usepackage{parskip}
  }{
    \setlength{\parindent}{0pt}
    \setlength{\parskip}{6pt plus 2pt minus 1pt}}
}{
  \KOMAoptions{parskip=half}}
\makeatother
\usepackage{xcolor}
\usepackage[top=30mm,left=20mm,heightrounded]{geometry}
\makeatletter
\ifx\paragraph\undefined\else
  \let\oldparagraph\paragraph
  \renewcommand{\paragraph}{
    \@ifstar
      \xxxParagraphStar
      \xxxParagraphNoStar
  }
  \newcommand{\xxxParagraphStar}[1]{\oldparagraph*{#1}\mbox{}}
  \newcommand{\xxxParagraphNoStar}[1]{\oldparagraph{#1}\mbox{}}
\fi
\ifx\subparagraph\undefined\else
  \let\oldsubparagraph\subparagraph
  \renewcommand{\subparagraph}{
    \@ifstar
      \xxxSubParagraphStar
      \xxxSubParagraphNoStar
  }
  \newcommand{\xxxSubParagraphStar}[1]{\oldsubparagraph*{#1}\mbox{}}
  \newcommand{\xxxSubParagraphNoStar}[1]{\oldsubparagraph{#1}\mbox{}}
\fi
\makeatother

\usepackage{longtable,booktabs,array}
\usepackage{calc} 
\usepackage{etoolbox}
\makeatletter
\patchcmd\longtable{\par}{\if@noskipsec\mbox{}\fi\par}{}{}
\makeatother
\IfFileExists{footnotehyper.sty}{\usepackage{footnotehyper}}{\usepackage{footnote}}
\makesavenoteenv{longtable}
\usepackage{graphicx}
\makeatletter
\newsavebox\pandoc@box
\newcommand*\pandocbounded[1]{
  \sbox\pandoc@box{#1}%
  \Gscale@div\@tempa{\textheight}{\dimexpr\ht\pandoc@box+\dp\pandoc@box\relax}%
  \Gscale@div\@tempb{\linewidth}{\wd\pandoc@box}%
  \ifdim\@tempb\p@<\@tempa\p@\let\@tempa\@tempb\fi
  \ifdim\@tempa\p@<\p@\scalebox{\@tempa}{\usebox\pandoc@box}%
  \else\usebox{\pandoc@box}%
  \fi%
}
\def\fps@figure{htbp}
\makeatother
\NewDocumentCommand\citeproctext{}{}

\makeatletter
 \let\@cite@ofmt\@firstofone
 \def\@biblabel#1{}
 \def\@cite#1#2{{#1\if@tempswa , #2\fi}}
\makeatother
\newlength{\cslhangindent}
\newlength{\csllabelwidth}
\newenvironment{CSLReferences}[2] 
 {\begin{list}{}{%
  \setlength{\itemindent}{0pt}
  \setlength{\leftmargin}{0pt}
  \setlength{\parsep}{0pt}
  \ifodd #1
   \setlength{\leftmargin}{\cslhangindent}
   \setlength{\itemindent}{-1\cslhangindent}
  \fi
  \setlength{\itemsep}{#2\baselineskip}}}
 {\end{list}}
\usepackage{calc}

\usepackage[noblocks]{authblk}

\makeatletter
\@ifpackageloaded{caption}{}{\usepackage{caption}}
\AtBeginDocument{%
\ifdefined\contentsname
  \renewcommand*\contentsname{Table of contents}
\else
  \newcommand\contentsname{Table of contents}
\fi
\ifdefined\listfigurename
  \renewcommand*\listfigurename{List of Figures}
\else
  \newcommand\listfigurename{List of Figures}
\fi
\ifdefined\listtablename
  \renewcommand*\listtablename{List of Tables}
\else
  \newcommand\listtablename{List of Tables}
\fi
\ifdefined\figurename
  \renewcommand*\figurename{Figure}
\else
  \newcommand\figurename{Figure}
\fi
\ifdefined\tablename
  \renewcommand*\tablename{Table}
\else
  \newcommand\tablename{Table}
\fi
}
\@ifpackageloaded{float}{}{\usepackage{float}}
\floatstyle{ruled}
\@ifundefined{c@chapter}{\newfloat{codelisting}{h}{lop}}{\newfloat{codelisting}{h}{lop}[chapter]}
\floatname{codelisting}{Listing}

\makeatother
\makeatletter
\@ifpackageloaded{caption}{}{\usepackage{caption}}
\@ifpackageloaded{subcaption}{}{\usepackage{subcaption}}
\makeatother

\usepackage{bookmark}

\IfFileExists{xurl.sty}{\usepackage{xurl}}{} 
\hypersetup{
  pdftitle={Breaching 1.5°C: Give me the odds},
  pdfkeywords={Paris Agreement, global warming, climate
change, 1.5°C, probability},
  colorlinks=true,
  linkcolor={blue},
  filecolor={Maroon},
  citecolor={Blue},
  urlcolor={Blue},
  pdfcreator={LaTeX via pandoc}}

\title{Breaching 1.5°C: Give me the odds\footnote{Corresponding author: \href{mailto: eduardo@math.aau.dk}{eduardo@math.aau.dk}}}

  \author{J. Eduardo Vera-Valdés}
            \affil{%
                  Aalborg University
              }
        \author{Olivia Kvist}
            \affil{%
                  Aalborg University
              }
      
\date{2024-12-17}
\begin{document}
\maketitle
\begin{abstract}
Climate change communication is crucial to raising awareness and
motivating action. In the context of breaching the limits set out by the
Paris Agreement, we argue that climate scientists should move away from
point estimates and towards reporting probabilities. Reporting
probabilities will provide policymakers with a range of possible
outcomes and will allow them to make informed timely decisions. To
achieve this goal, we propose a method to calculate the probability of
breaching the limits set out by the Paris Agreement. The method can be
summarized as predicting future temperatures under different scenarios
and calculating the number of possible outcomes that breach the limits
as a proportion of the total number of outcomes. The probabilities can
be computed for different time horizons and can be updated as new data
become available. As an illustration, we performed a simulation study to
investigate the probability of breaching the limits in a statistical
model. Our results show that the probability of breaching the 1.5°C
limit is already greater than zero for 2024. Moreover, the probability
of breaching the limit is greater than 99\% by 2042 if no action is
taken to reduce greenhouse gas emissions. Our methodology is simple to
implement and can easily be extended to more complex models of the
climate system. We encourage climate model developers to include the
probabilities of breaching the limits in their reports.
\end{abstract}

\subsection{The 1.5°C limit}\label{the-1.5c-limit}

The goals of the Paris Agreement (PA) have recently gained renewed media
attention due to observed temperature anomalies that exceeded 1.5°C
above preindustrial levels for 12 consecutive months according to
Copernicus Climate Change Service (2024a). The importance of the 1.5°C
threshold is that it was established in the PA as a limit to avoid the
most severe consequences of climate change. Formally, the PA aims to
limit global warming to well below 2°C above pre-industrial levels and
to pursue efforts to limit the temperature increase to 1.5°C.

An obstacle in assessing the success or failure of the PA is the lack of
a clear definition of when temperature limits are breached (Betts et al.
2023). The definition of when the limits are breached is crucial for
both scientific and political reasons.

\textbf{If we defined the breaching of 1.5°C as the mean temperature for
a year being above that limit, it has already been breached.}

However, to avoid short-term fluctuations, the Sixth Assessment Report
of Working Group I of the Intergovernmental Panel on Climate Change
(IPCC) proposes to use a 20-year average temperature rise to determine
when the limit is exceeded (IPCC 2021). The question remained on when
inside that 20-year period the limit is breached.

Betts et al. (2023) argue that defining the breach of the 1.5°C limit as
the last year in a 20-year period where the global mean temperature is
above that limit delays the conclusion of a breach by a decade. They
propose using the midpoint of the 20-year period as the year when the
limit is breached. Thus, computing when the threshold will be breached
entails averaging several years of observed temperature rise with a
forecast of the following years up to the 20-year period. We extend this
methodology to provide the probability of breaching the 1.5°C and 2°C
limits with the aim of improving the communication of climate change.

\subsection{Improving communication of climate
change}\label{improving-communication-of-climate-change}

One of the main challenges in communicating climate change is the
complexity of the topic. This complexity makes it difficult to
communicate the issue in a way that is easily understandable to the
general public. In the context of breaching the limits set out by the
PA, communication is crucial. The issue can become highly politicized if
not communicated effectively. The public and policymakers need timely
information about the urgency of the situation and the consequences of
inaction.

One of the first steps in improving communication is to provide data in
a clear and understandable way. Datasets report temperature anomalies as
the difference between the observed temperature and the average
temperature for a reference period (GISTEMP 2020; Morice et al. 2021; R.
A. Rohde and Hausfather 2020). Even though the PA states that the
reference period should be pre-industrial levels, the datasets typically
use a more recent reference period. For example, the HadCRUT5 dataset
uses the 1961-1990 average temperature as the reference period.

\begin{figure}[H]

\centering{

\includegraphics[scale=0.55]{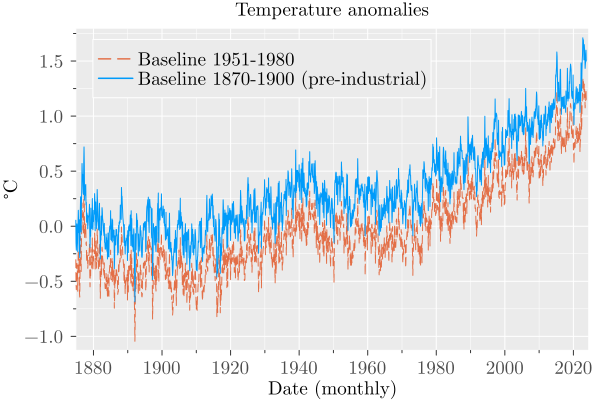}

}

\caption{\label{fig-last}Temperature anomalies (°C) in the HadCRUT5
dataset (Morice et al. 2021). The dashed line presents the data
according to the baseline period in HadCRUT5 (1951-1980). The solid line
represents the temperature anomalies with the pre-industrial baseline
period (1870-1900).}

\end{figure}%


Figure~\ref{fig-last} shows temperature anomalies as reported by the
HadCRUT5 dataset. The figure shows that if we use the 1961-1990 average
temperature as the reference period, as presented in the dataset, the
temperature anomalies have not breached the 1.5°C limit yet. However, if
we use the pre-industrial levels as the reference period, as indicated
in the PA, the limit has already been breached several times. This
mismatch between the reference period used in the datasets and the
reference period in the PA can lead to misunderstandings and
misinterpretations. A sceptic reading a news article reporting
temperature anomalies breaching the 1.5°C limit above pre-industrial
levels can easily download and plot the data getting the impression that
the headline is an exaggeration if they are not aware of the reference
period used.

\textbf{All datasets should use the same reference period based on the
pre-industrial levels.} This will help to avoid confusion and to make it
easier to compare the data. However, for historical reasons, data
providers should also report temperature anomalies relative to their
original reference period. This will help maintain compatibility with
previous reports and models trained on the original data.

\subsection{Predictions for the breaching of the PA
limits}\label{predictions-for-the-breaching-of-the-pa-limits}

It should be stressed in any report that determining when the 1.5°C
limit will be breached requires forecasting future temperatures.
Forecasts can take many forms. The most common are physical models that
simulate the climate system {[}see e.g.; Nath et al. (2022); Eyring et
al. (2016); Held et al. (2019); Collins, Tett, and Cooper (2001); Orbe
et al. (2020){]}. Physics-based models are computationally expensive and
require high-performance computing. Hence, reduced-complexity models
have been developed. These models are based on statistical methods and
are trained on historical data of different climate variables {[}see
e.g.; Meinshausen, Raper, and Wigley (2011); Bennedsen, Hillebrand, and
Koopman (2024){]}.

Regardless of the method used to predict future temperatures, forecasts
are uncertain. The climate system is complex and chaotic. This
complexity is reflected in the confidence intervals associated with the
forecasts. For example, the IPCC provides a range of possible outcomes
for future temperatures. However, the uncertainty in the forecasts is
not communicated effectively when discussing breaching the limits set
out by the PA.

The media has recently reported new estimates on when the 1.5°C limit
will be breached (Copernicus Climate Change Service 2024b; R. Rohde
2024). However, these estimates are often presented as point estimates
without confidence intervals or without a clear description of the
methodology used to make the predictions. In the current political
environment, it is crucial to communicate the uncertainty in the
predictions.

Recent point estimates of when the 1.5°C limit will be breached can be
counterproductive if not accompanied by probability estimates. In case
the limit is not breached in precisely the year predicted, it can give
climate change deniers an argument to dismiss scientific evidence. In
the past, extreme winters have been used as an argument against global
warming due to the misunderstanding of the difference between weather
and climate. Where weather refers to something more local and only
observed over short-time periods, climate is more long-termed.
\textbf{The distinction between weather and climate must be clear in any
communication to avoid misrepresentation of the results.}

\subsection{A new methodology to measure when we will breach the limit
of
1.5°C}\label{a-new-methodology-to-measure-when-we-will-breach-the-limit-of-1.5c}

We propose a way to communicate the uncertainty in the predictions of
when the limits set at the PA will be breached. The methodology builds
on the proposal by Betts et al. (2023) to use a 20-year average
temperature rise centered around a particular year. The 20-year average
is then compared with the 1.5°C and 2°C limits. We use models to produce
multiple scenarios of future temperature rise and compute the number of
scenarios that breach the limits as a proportion of the total number of
scenarios. The probabilities can be computed for different time horizons
and can be updated as new data become available. Moreover, the
methodology can be easily applied for different climate models and
datasets.

There are already several examples of how probabilities can be used to
communicate climate change effectively {[}see e.g.; IPCC (2021); Wigley
and Raper (2001); S. H. Schneider (2001); S. H. Schneider and
Mastrandrea (2005); T. Schneider et al. (2023){]}. \textbf{By reporting
probabilities, we can communicate the uncertainty in the predictions and
provide policymakers with a range of possible outcomes.} This will allow
policymakers to make more informed decisions on taking action to reduce
greenhouse gas emissions. Reporting in 2024 a probability of 50\% that
the limit will be breached in 2030 will give an indication of the
urgency of the situation. The probability distribution will also reflect
how the odds of avoiding the breach decrease over time if no action is
taken. This will provide a clear picture of the consequences of delaying
action.

To illustrate our methodology, we developed a simulation study. We
simulate multiple scenarios of future temperature rise and calculate the
probability of breaching the 1.5°C and 2°C limits. The simulation study
is presented next.

\subsection{A statistical model to predict future
temperatures}\label{a-statistical-model-to-predict-future-temperatures}

\textbf{Data.} The data used in this paper is the global mean
temperature anomaly of the HadCRUT5 dataset computed by the Met Office
Hadley Centre Morice et al. (2021). The data are reported as the
difference between the observed temperature and the 1961-1990 average
temperature and are available from 1850. We first convert the data to
anomalies compared to pre-industrial levels. The pre-industrial levels
are defined as the average temperature from the earliest available data
up to 1900. The data is presented in Figure~\ref{fig-last}.

HadCRUT5 provides 200 realizations to account for the uncertainty in the
data. We use all realizations to fit the models and produce multiple
scenarios of future temperature rise. This allows us to account for the
uncertainty in the data and to provide a range of possible outcomes. We
fit the models to each realization separately and produce five different
scenarios of future temperatures for each realization. This gives us a
total of 1000 scenarios of future temperatures. The methodology can be
easily extended to include more realizations and scenarios.

\textbf{Modeling scheme.} Our modeling scheme consists of three
components: a trend specification, an El Niño Southern Oscillation
(ENSO) model, and a long-range dependent error term. We provide a brief
overview of the models. Further technical details on the models are
presented in the supplementary material in the appendix, and the code
used to perform the simulation study is available in a
\href{https://everval.github.io/Odds-of-breaching-1.5C/Notebook-HadCRUT-preview.html}{Jupyter notebook in the supplementary
material}.

We consider three trend specifications for modeling the global mean
temperature anomaly: a linear trend model, a quadratic trend model, and
a linear trend that allows for a break. The models are estimated on the
historical temperature data. The best model is selected on the basis of
the Akaike Information Criterion (AIC) and Bayesian Information
Criterion (BIC) (Akaike 1974; Schwarz 1978). For each realization, the
model with the lowest AIC and BIC is considered the best model and is
used to predict future temperatures.

Furthermore, we control for the El Niño effect as it is known to have an
effect on the global mean temperature anomaly (Thirumalai et al. 2017;
Jiang et al. 2024). To control for the El Niño effect we include the
Oceanic Niño Index (ONI) as a covariate in the models. The ONI is an
indicator for monitoring the ENSO. El Niño conditions are present when
the ONI is +0.5 or higher. Oceanic La Niña conditions exist when the ONI
is -0.5 or lower.

For forecasting purposes, we fit a Markov-switching model to the ONI
data to predict future values (Hamilton 1989, 1990). The motivation for
using a Markov-switching model is that the ONI data naturally exhibit
regime changes over time. The number of states in the Markov-switching
model is 7, which is selected on the basis of the AIC and BIC. The seven
states correspond to the different phases of the ENSO cycle, ranging
from very strong El Niño, strong El Niño, moderate El Niño, neutral,
moderate La Niña, strong La Niña, to very strong La Niña.

Finally, our modeling scheme allows for the error term to have
long-range dependence. Long-range dependence has its origin in the
analysis of climate data (Hurst 1956). Temperature data are known to
have long-range dependence, which means that the error terms are
correlated over long periods (Bloomfield and Nychka 1992; Bloomfield
1992; J. Eduardo Vera-Valdés 2021). The long-range dependence parameter
is estimated using the exact local Whittle method (Shimotsu and Phillips
2005).

\textbf{Model validation.} We obtain the prediction intervals for
temperature anomalies using our modeling scheme fitted to data up to
November 2016, the month when the PA entered into force. All HadCRUT5
realizations are considered. The results, presented in the supplementary
material, show that our models provide adequate coverage of the observed
temperature anomalies up to the present day. We take this as validation
of our modeling scheme.

\begin{figure}[H]

\centering{

\includegraphics[scale=0.55]{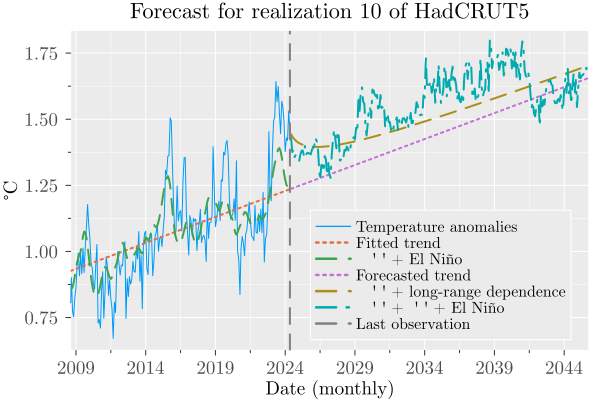}

}

\caption{\label{fig-models-forecasts}Forecast of temperature anomalies
for realization 10 of the HadCRUT5 dataset. The forecast is based on the
broken trend model with long-range dependence and El Niño as an
exogenous variable. A simulated El Niño series using a Markov-switching
model with 7 states was used to generate the forecast.}

\end{figure}

\textbf{Model fitting.} As an illustration, we present a fitted model
and its forecast for realization 10 of the HadCRUT5 dataset. Realization
10 is chosen arbitrarily. The model is fitted to the data up to the last
observation. The model is then used to forecast future temperatures. The
results are presented in Figure~\ref{fig-models-forecasts}.

Figure~\ref{fig-models-forecasts} highlights the different components of
the model: the trend, the long-range dependence, and the El Niño effect.

The trend component captures the long-term increase in the temperature
anomaly, all other things being equal. The long-range dependence
captures the persistence of the temperature anomaly over time. Given
that recent temperatures are high, the long-range dependence in the data
implies that future temperatures are likely to remain high. This
directly affects the forecasted temperature and the probability of
breaching the limits. Finally, the El Niño effect captures the
short-term fluctuations in the temperature anomaly. The forecasted
temperature anomaly is the sum of the trend, the long-range dependence,
and the El Niño effect.

\textbf{Breaching the limits.} For each simulated path, we calculate the
average temperature for 20 years using a moving average. We began the
process in 2004 to obtain a 20-year average temperature rise centered
around 2014 and with an end point in the current year. The moving
average is then calculated for each month. We repeat this process until
the end of the forecast period. We then find the first month where the
20-year average temperature rise breaches the 1.5°C and 2°C limits.

\begin{figure}[H]

\centering{

\includegraphics[scale=0.55]{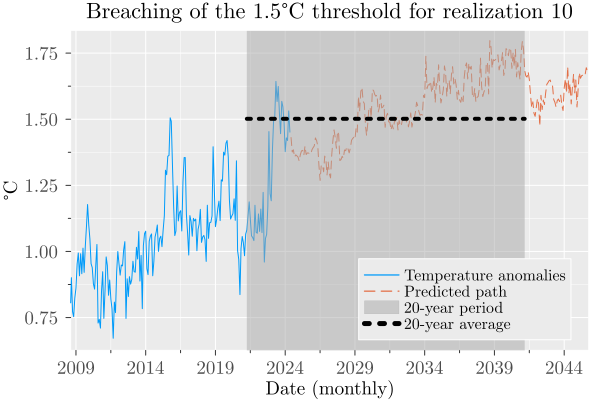}

}

\caption{\label{fig-breaching-static}Breaching of the 1.5°C threshold
for realization 10 of the HadCRUT5 dataset. The figure shows the
temperature anomalies and the forecasted path for the next several
months. The 20-year period is highlighted in gray, and the 20-year
average is shown as a black dashed line.}

\end{figure}%


Figure~\ref{fig-breaching-static} shows that the 20-year average
temperature for the simulated path of realization 10 first breaches the
1.5°C limit in July of 2031. The gray box indicates the 20-year period
used to calculate the average temperature rise, while the black dashed
line indicates the 20-year average temperature.

The month in which the limit is breached for this path is highly
dependent on the El Niño effect. Hence, we conduct a simulation study to
estimate the probability of breaching the limits.

\subsection{Simulation study}\label{simulation-study}

Using the modeling scheme described above, we detail a way to compute
the probability of breaching the limits set out by the PA using a
simulation study. The use of Monte Carlo methods, as the one used in
this simulation study, is a common approach to estimate probabilities in
complex systems, and it is pursued by the IPCC (Abel, Eggleston, and
Pullus 2002). The simulation study has two main steps.

\begin{figure}[H]

\centering{

\includegraphics[scale=0.55]{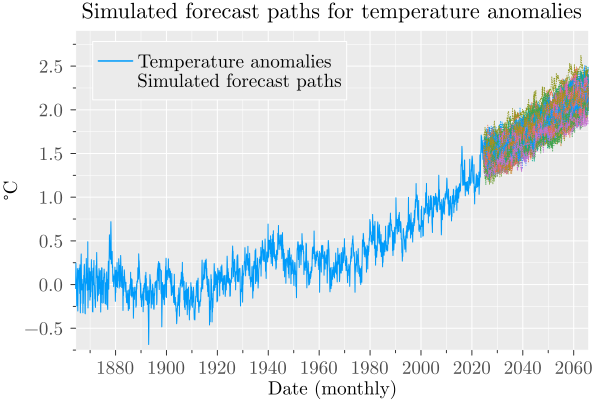}

}

\caption{\label{fig-paths}Simulated forecast paths for HadCRUT5
temperature anomalies. One hundred paths of a total of 1000 paths are
shown to ease visualization. The forecasts are based on the best-fitting
model for each realization, with El Niño as an exogenous variable. For
ease of visualization, the mean of all temperature anomaly realizations
is shown as a solid line.}

\end{figure}

First, we forecast the global mean temperature anomaly using the best
model selected using the information criteria. For each realization of
the HadCRUT5 dataset, we simulate 5 different scenarios of future
temperature rise by simulating different paths for El Niño effect. This
gives us a total of 1000 scenarios of future temperatures.
Figure~\ref{fig-paths} shows the simulated temperature anomalies for a
subset of the realizations to simplify visualization and plot rendering.

\begin{figure}[H]

\centering{

\includegraphics[scale=0.55]{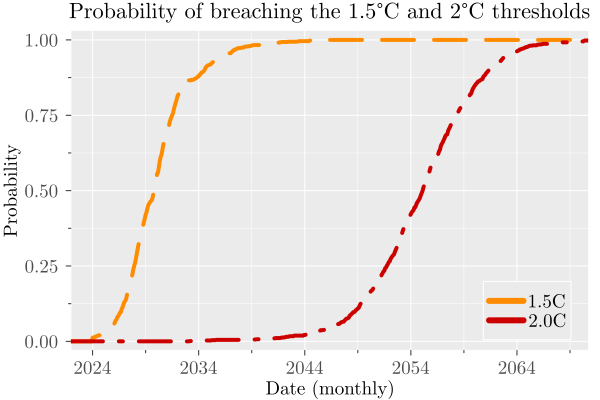}

}

\caption{\label{fig-simulation-dist}Proportion of scenarios that breach
the 1.5°C and 2°C thresholds for the HadCRUT5 temperature anomalies for
each month. The figure considers 1000 scenarios, each based on the
best-fitting model for each realization, with five simulations for El
Niño as an exogenous variable each.}

\end{figure}

In a second step, we calculate the 20-year moving average centered
around a particular month for each simulated path. We repeat this
process for all simulated paths and recover the ratio of paths that
breach the 1.5°C and 2°C limits each month to the total number of paths.
We then plot this proportion of paths that crossed either threshold to
obtain an estimate of the probability of breaching the limits.
Figure~\ref{fig-simulation-dist} presents the results of the simulation
study.

\textbf{The simulation study considered here shows that the probability
of breaching the 1.5°C limit is already greater than zero for 2024.}

This means that there is at least one scenario in which the 20-year
average temperature rise breaches the 1.5°C limit in September 2024.
Moreover, note that there is a rapid increase in the probability of
breaching the 1.5°C limit after 2030. The probability of breaching the
limit is already greater than 50\% by July 2030. This is in line with
recent predictions that the goal will likely be breached in the second
half of the 2030 decade (Copernicus Climate Change Service 2024b; R.
Rohde 2024). Our simulation study provides an estimate of the monthly
probabilities of breaching the goals. They show that the probability of
breaching the 1.5°C limit is greater than 99\% by the end of 2042 if no
action is taken to reduce greenhouse gas emissions.

Regarding the 2°C limit, our simulation study finds that the probability
of breach already starts increasing above zero by the 2030 decade. In
general, the simulation study highlights that climate change mitigation
policies should be implemented as soon as possible to avoid breaching
the limits set by the PA.

\subsection{Discussion and further
work}\label{discussion-and-further-work}

We have presented a new way to communicate when we will breach the
temperature limits set out by the PA. Our methodology is simple to
implement. It requires predicting future temperatures under different
scenarios and calculating the number of possible outcomes that breach
the limits as a proportion of the total number of outcomes. The
probabilities can be computed for different time horizons and datasets
and can be updated as new data becomes available. Additional simulation
exercises considering alternative datasets and sub-samples of
realizations are presented in the supplementary material.

We have illustrated the methodology in a simulation study. The
simulation study is based on statistical models trained on historical
temperature data to predict future temperatures. Our results are based
on the assumption that no structural changes will occur in the future.
In that sense, our results could be interpreted as a scenario in which
no action is taken to reduce greenhouse gas emissions from the current
levels.

The methodology can be easily extended to include different scenarios of
future emissions and more complex models of the climate system. Climate
models such as MAGICC already provide a range of possible outcomes for
future temperatures; our methodology can be easily applied to these
models. We encourage climate model developers to include the
probabilities of breaching the limits in their reports.

\subsubsection{Reproducibility}\label{reproducibility}

The code used to perform the simulation study is available in a
\href{https://everval.github.io/Odds-of-breaching-1.5C/Notebook-HadCRUT-preview.html}{Jupyter notebook in the supplementary
material}. The code is written in Julia (Bezanson et al. 2017). The
Julia programming language is a high-level and high-performance language
for technical computing. Additional packages used in the simulation
study are the \texttt{DataFrames.jl} package for data manipulation
(Bouchet-Valat and Kamiński 2023), the \texttt{MarSwitching.jl} package
for Markov-switching models (Dadej 2024), the \texttt{LongMemory.jl}
package for long-range dependent models (J. E. Vera-Valdés 2024), the
\texttt{CSV.jl} package to read and write CSV files (Quinn et al. 2024),
and the \texttt{Plots.jl} package for plotting (Breloff 2024).

The code is well documented and includes comments to explain the
different steps of the simulation study. The code is open-source and can
be freely used and modified. We encourage other researchers to use the
code to reproduce our results and to extend the methodology to other
datasets and models.

\subsection*{References}\label{references}
\addcontentsline{toc}{subsection}{References}

\phantomsection\label{refs}

\section{Supplementary material}\label{sec-supplementary}

The supplementary material contains additional information on the models
used in the simulation study. The components of the models are described
in detail.

\subsection{Trend models}\label{trend-models}

We consider three trend specifications for modeling the global mean
temperature anomaly: a linear trend model, a quadratic trend model, and
a linear trend allowing for a break. The models are given by:

\begin{itemize}
\item
  Linear Trend:
  \(y_t = \beta_0 + \beta_1 t + \gamma ONI_t + \epsilon_t,\)
\item
  Quadratic Trend:
  \(y_t = \beta_0 + \beta_1 t + \beta_2 t^2 + \gamma ONI_t + \epsilon_t,\)
\item
  Trend with Break:
  \(y_t = \beta_0 + \beta_1 t + \beta_2 I_{t > t_0} + \gamma ONI_t + \epsilon_t.\)
\end{itemize}

Above, \(y_t\) is the global mean temperature anomaly at time \(t\),
\(\beta_0\), \(\beta_1\), and \(\beta_2\) are the trend coefficients,
\(\gamma\) is the coefficient of the El Niño effect, \(ONI_t\) is the
variable that models the El Niño events, and \(\epsilon_t\) is the error
term. As described in the following, the error term is assumed to have
long-range dependence. The variable \(I_{t > t_0}\) is an indicator
variable that takes the value 1 if \(t > t_0\) and 0 otherwise. The
break point \(t_0\) is estimated from the data.

The models are estimated on the historical temperature data. The best
model is selected based on the Akaike Information Criterion (AIC) and
Bayesian Information Criterion (BIC) (Akaike 1974; Schwarz 1978). For
each realization, the model with the lowest AIC and BIC is considered
the best model and is used to predict future temperatures.

For example, the AIC and BIC for the trend models fitted to realization
10 are presented in the table below.

\begin{longtable}[]{@{}lll@{}}
\toprule\noalign{}
Model & AIC & BIC \\
\midrule\noalign{}
\endhead
\bottomrule\noalign{}
\endlastfoot
Linear Trend & -5613.2 & -5596.64 \\
Quadratic Trend & -6551.17 & -6529.09 \\
Trend with Break & -6627.33 & -6605.25 \\
\end{longtable}

The estimated coefficient confidence intervals are used to simulate
future values of the temperature anomaly. The confidence intervals are
obtained from the coefficients' (asymptotic) distribution. Under
normally distributed error term, the coefficient estimators are normally
distributed with mean and variance given by the following formula:

\[\hat{\beta} \sim N(\beta, \sigma^2(X'X)^{-1}),\]

where \(\hat{\beta}\) are the estimates, \(\beta\) are the true
coefficients, \(\sigma^2\) is the variance of the error term, and \(X\)
is the design matrix. In case of non-normal error term, the coefficient
estimators are asymptotically normal using the central limit theorem
under mild conditions (Wooldridge 2010).

\subsection{El Niño Southern Oscillation (ENSO)
model}\label{el-niuxf1o-southern-oscillation-enso-model}

El Niño Southern Oscillation (ENSO) is a natural climate phenomenon that
influences global temperature. It is characterized by periodic warming
of sea surface temperatures in the central and eastern equatorial
Pacific Ocean. It is observed every 2-7 years and can last from 9 months
to 2 years.

Modeling the El Niño effect is crucial for predicting future
temperatures. To control for the El Niño effect, we include the Oceanic
Niño Index (ONI) as a covariate in the models as described above. The
ONI is an indicator for monitoring the ENSO. El Niño conditions are
present when the ONI is +0.5 or higher. Oceanic La Niña conditions exist
when the ONI is -0.5 or lower.

One complication with the El Niño effect is that it is difficult to
predict. The El Niño events are highly variable and can have different
intensities. The El Niño effect can also interact with other climate
phenomena, such as the Indian Ocean Dipole and the Madden-Julian
Oscillation. This makes it challenging to model the El Niño effect
accurately {[}see e.g.; Thirumalai et al. (2024); Ham, Kim, and Luo
(2019); L'Heureux et al. (2020); Hassanibesheli, Kurths, and Boers
(2022){]}. In this study, we use a simple model to capture the El Niño
effect. The model is based on the historical ONI data and is used to
simulate future ONI values.

The dynamics of the ONI are modeled using a Markov-switching model
(Hamilton 1989). The Markov-switching model is a regime-switching model
that allows for the presence of different regimes in the data. The model
is given by:

\[ONI_t = \beta_{j} + \epsilon_{j,t},\]

where \(\beta_{j}\) is the coefficient for the \(j\)-th regime, and
\(\epsilon_{j,t}\) is the error term with variance \(\sigma^2_j\). A
latent state at time \(t\), \(s_t\), indicates the regime. The dynamics
of \(s_t\) are governed by a Markov process:
\[ Pr(s_t = j | s_{t-1} = i, s_{t-2}, \cdots, s_1) = Pr(s_t = j | s_{t-1} = i) = p_{ij},\]
where \(p_{ij}\) is the transition probability from state \(i\) to
\(j\).

Note that the probability distribution of \(s_t\) given the entire path
\(\left\{s_{t-1}, s_{t-2}, \cdots, s_1\right\}\) depends only on the
most recent state \(s_{t-1}\).

In historical data, the effect can be estimated using maximum likelihood
estimation and the expectation-maximization algorithm (Hamilton 1990).
For forecasting, the effect is simulated using a stochastic process
taking into account the probability of each regime.

To determine the number of regimes, we use the AIC and BIC. We consider
a range of possible regimes and select the number of regimes that
minimize the AIC and BIC. The following table shows the AIC and BIC for
the ONI data. Only odd numbers of regimes are considered to ensure that
the model includes both El Niño and La Niña events and neutral
conditions.

\begin{longtable}[]{@{}lll@{}}
\toprule\noalign{}
Regimes & AIC & BIC \\
\midrule\noalign{}
\endhead
\bottomrule\noalign{}
\endlastfoot
3-regimes & 2438 & 2504 \\
5-regimes & 2342 & 2507 \\
7-regimes & 1394 & 1703 \\
\end{longtable}

Hence, the number of states in the Markov-switching model is 7. The 7
states are chosen to correspond to the different phases of the ENSO
cycle ranging from very strong El Niño, strong El Niño, moderate El
Niño, neutral, moderate La Niña, strong La Niña, to very strong La Niña.

\subsection{Long-range dependent error
term}\label{long-range-dependent-error-term}

Long-range dependent models imply that past values of the series have a
long-lasting effect on the current value. It describes the tendency for
successive values to remain close to each other or to be dependent.
Interestingly, the notion of long-range dependence originated in the
analysis related to climate data in the pioneering work of Hurst (1956)
on the Nile River minima. Hurst determined that a dam built to control
river flow should be designed to withstand the worst-case scenario. The
worst-case scenario was determined by the long-range dependence in the
data. Years with high minima were likely to be followed by years with
high minima. This phenomenon is known as the Joseph effect. This is due
to Joseph's interpretation in the Old Testament of Pharaoh's dream,
which predicted that seven years of plenty would be followed by seven
years of famine.

A long-range dependent model can be written as:
\[y_t = \sum_{j=1}^\infty \phi_j y_{t-j} + \epsilon_t,\] where
\(\epsilon_t\) is an i.i.d. process. The coefficients \(\phi_j\) decay
hyperbolically (slowly) to zero as \(j\) increases. In contrast, the
coefficients of standard models decay exponentially to zero.

The temperature series exhibit long-range dependence. In the context of
breaching the limits set out by the PA, the long-range dependence in the
data is crucial since it affects the forecasted temperature rise.

One likely explanation behind the presence of long-range dependence in
the data is aggregation (Clive W. J. Granger 1980; Zaffaroni 2004;
Haldrup and Vera-Valdés 2017). The global mean temperature anomaly is an
aggregate of temperature data from different regions. The aggregation
process can lead to long-range dependence in the data. To account for
this property, we model the error term in the trend models as a
long-range dependent process.

We used the exact local Whittle estimator to estimate the long-range
dependence in the data (Shimotsu and Phillips 2005). The exact local
Whittle estimator is a semi-parametric estimator that estimates the
long-range dependence parameter by maximizing the modified Whittle
likelihood function originally proposed by Künsch (1987).

The exact local Whittle estimator minimizes the function given by:
\[R(d) = \log\left(\frac{1}{m}\sum_{k=1}^{m}I_{\Delta^d}(\lambda_k)\right)-\frac{2d}{m}\sum_{k=1}^{m}\log(\lambda_k),\]

where \(I_{\Delta^d}(\lambda_k)\) is the periodogram of \((1-L)^d x_t\),
where \((1-L)^d\) is the fractional difference operator (C. W. J.
Granger and Joyeux 1980; Hosking 1981), \(\lambda_{k} = e^{i2\pi k /T}\)
are the Fourier frequencies, and \(m\) is the bandwidth parameter.

The exact local Whittle estimator is consistent and asymptotically
normal. The long-range dependence parameter is estimated for each
realization separately. The estimated parameter is then used to simulate
the error term in the models.

\subsection{How has the probabilities changed since the Paris
Agreement}\label{how-has-the-probabilities-changed-since-the-paris-agreement}

\begin{figure}[H]

\centering{

\includegraphics[scale=0.55]{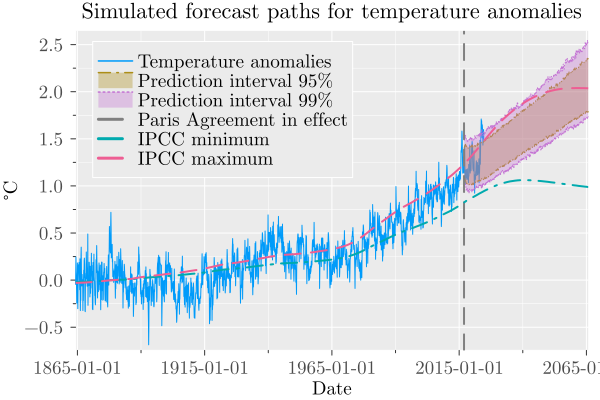}

}

\caption{\label{fig-paths-covprepa}Simulated forecast paths for HadCRUT5
temperature anomalies. The 95\% and 99\% prediction intervals are shown
as shaded areas. The IPCC projections for the minimum and maximum
temperature anomalies are shown as dashed lines (Allen et al. 2018).}

\end{figure}

As model validation, Figure~\ref{fig-paths-covprepa} presents the
prediction intervals for temperature anomalies for the modeling scheme
described above starting in November 2016, the month when the PA entered
into force. The results using the data up to the PA are
presented in the \href{https://everval.github.io/Odds-of-breaching-1.5C/Notebook-HadCRUT-BeforePA-preview.html}{supplementary Jupyter
notebook}.

The prediction intervals are based on the historical data up to the
start of the PA and the models fitted to the data. The prediction
intervals are used to assess the uncertainty in the forecasts. In
general, the prediction intervals provide adequate coverage of the
observed temperature anomalies. However, note that recent high
temperatures fall outside the 99\% prediction intervals. This further
signals the abnormality of the recent temperature observations. Several
theories have been proposed to explain recent high temperatures,
including decreased cloud coverage and international shipping regulatory
changes (Goessling, Rackow, and Jung; Quaglia and Visioni 2024).
Regardless of the cause, the high temperatures highlight the urgency of
the situation.

In contrast, the figure presents the temperature projections from the
summary for policymakers of the IPCC Special Report: Global Warming of
1.5°C (Allen et al. 2018). The paths show the projected temperature
evolution according to the IPCC if CO$_2$ emission gradually decrease
to zero by 2055, while other greenhouse gas levels stop changing after
2030. The figure shows that recent temperatures are outside the IPCC
projections. Hence, the IPCC projections coverage is lacking, and the
projections are likely to be too optimistic.

\begin{figure}[H]

\centering{

\includegraphics[scale=0.55]{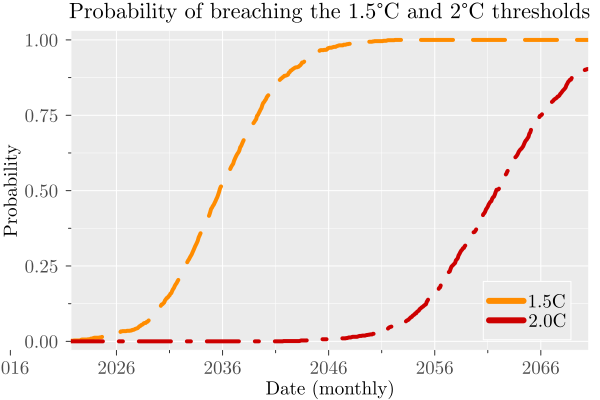}

}

\caption{\label{fig-dist-prepa}Proportion of scenarios that breach the
1.5°C and 2°C thresholds for the HadCRUT5 temperature anomalies for each
month at the start of the Paris Agreement. The figure considers 1000
scenarios, each based on the best-fitting model for each realization,
with five simulations for El Niño as an exogenous variable each.}

\end{figure}

Furthermore, Figure~\ref{fig-dist-prepa} presents the probabilities of
breaching the 1.5°C and 2°C limits at the start of the PA.

The figure allows us to assess how the probability of breaching the
limits has changed since the PA. At the start of the PA, the probability
of breaching the 1.5°C limit with a probability of 99\% was not
encountered until 2051. The probability of breaching the 2°C limit at a
probability of 99\% was not encountered in the forecast period ending in
2083. The results are related to the exercise of Copernicus Climate
Change Service (2023) on the time \emph{lost} since the PA considering a
point estimate, while we provide the probabilities of breaching the
limits. Probabilities have increased significantly since the PA, which
highlights that the urgency of the situation has increased since the PA.

\subsection{Alternative data sources}\label{alternative-data-sources}

The simulation study is based on the HadCRUT5 dataset. However, the
methodology can be easily extended to include other datasets. For
example, the GISTEMP dataset (GISTEMP 2020) provides an alternative
temperature anomalies data. The GISTEMP dataset is produced by the NASA
Goddard Institute for Space Studies and provides global temperature
anomalies data from 1880. The results using the GISTEMP dataset are
presented in the \href{https://everval.github.io/Odds-of-breaching-1.5C/Notebook-GISTEMP-preview.html}{supplementary Jupyter
notebook}.

They show that the probability of breaching the 1.5°C limit is already
greater than zero for May of 2027. Moreover, the probability of
breaching it is greater than 99\% by 2043. The results are in line with
the results obtained using the HadCRUT5 dataset. This shows that the
methodology can be easily extended to include other datasets.


\begin{CSLReferences}{1}{0}
\bibitem[\citeproctext]{ref-abel2002quantifying}
Abel, K, S Eggleston, and T Pullus. 2002. {``Quantifying Uncertainties
in Practice. IPCC Good Practice Guidance and Uncertainty Management in
National Greenhouse Gas Inventories.''}

\bibitem[\citeproctext]{ref-AIC1974}
Akaike, H. 1974. {``A New Look at the Statistical Model
Identification.''} \emph{IEEE Transactions on Automatic Control} 19 (6):
716--23. \url{https://doi.org/10.1109/TAC.1974.1100705}.

\bibitem[\citeproctext]{ref-IPCCSR15}
Allen, Myles, Opha Pauline Dube, William Solecki, Fernando
Aragón-Durand, Wolfgang Cramer, Stephen Humphreys, Mikiko Kainuma, et
al. 2018. {``Special Report: Global Warming of 1.5 c.''}
\emph{Intergovernmental Panel on Climate Change (IPCC)} 27: 677.

\bibitem[\citeproctext]{ref-Bennedsen2024}
Bennedsen, Mikkel, Eric Hillebrand, and Siem Jan Koopman. 2024. {``{A
Statistical Reduced Complexity Climate Model for Probabilistic Analyses
and Projections}.''} \emph{{arXiv}}.
\url{https://arxiv.org/abs/2407.04351}.

\bibitem[\citeproctext]{ref-BETTS2023}
Betts, Richard A., Stephen E. Belcher, Leon Hermanson, Albert Klein
Tank, Jason A. Lowe, Chris D. Jones, Colin P. Morice, Nick A. Rayner,
Adam A. Scaife, and Peter A. Stott. 2023. {``{Approaching
\(1.5^{\circ}\)C: how will we know we've reached this crucial warming
mark?}''} \emph{Nature} 624: 33--35.
\url{https://doi.org/10.1038/d41586-023-03775-z}.

\bibitem[\citeproctext]{ref-bezanson2017julia}
Bezanson, Jeff, Alan Edelman, Stefan Karpinski, and Viral B Shah. 2017.
{``Julia: A Fresh Approach to Numerical Computing.''} \emph{SIAM Review}
59 (1): 65--98. \url{https://doi.org/10.1137/141000671}.

\bibitem[\citeproctext]{ref-Bloomfield1992a}
Bloomfield, Peter. 1992. {``{Trends in global temperature}.''}
\emph{Climatic Change} 21 (1): 1--16.
\url{https://doi.org/10.1007/BF00143250}.

\bibitem[\citeproctext]{ref-Bloomfield1992}
Bloomfield, Peter, and Douglas Nychka. 1992. {``{Climate spectra and
detecting climate change}.''} \emph{Climatic Change} 21 (3): 275--87.
\url{https://doi.org/10.1007/BF00139727}.

\bibitem[\citeproctext]{ref-JSSv107i04}
Bouchet-Valat, Milan, and Bogumił Kamiński. 2023. {``DataFrames.jl:
Flexible and Fast Tabular Data in Julia.''} \emph{Journal of Statistical
Software} 107 (4): 1--32. \url{https://doi.org/10.18637/jss.v107.i04}.

\bibitem[\citeproctext]{ref-tom_breloff_2024_14094364}
Breloff, Tom. 2024. {``Plots.jl.''} Zenodo.
\url{https://doi.org/10.5281/zenodo.14094364}.

\bibitem[\citeproctext]{ref-HadCM3}
Collins, M., S. F. B. Tett, and C. Cooper. 2001. {``The Internal Climate
Variability of HadCM3, a Version of the Hadley Centre Coupled Model
Without Flux Adjustments.''} \emph{Climate Dynamics} 17 (1): 61--81.
\url{https://doi.org/10.1007/s003820000094}.

\bibitem[\citeproctext]{ref-CopernicusJanuary24}
Copernicus Climate Change Service. 2024a. {``{Surface air temperature
for January 2024}.''} {Copernicus}.
\url{https://climate.copernicus.eu/surface-air-temperature-january-2024}.

\bibitem[\citeproctext]{ref-C3SParis}
Copernicus Climate Change Service, C3S. 2023. {``We've {`Lost'} 19 Years
in the Battle Against Global Warming Since the Paris Agreement.''}
\emph{European Centre for Medium-Range Weather Forecasts}.
\url{https://climate.copernicus.eu/weve-lost-19-years-battle-against-global-warming-paris-agreement}.

\bibitem[\citeproctext]{ref-C3SApp}
---------. 2024b. {``C3S Global Temperature Trend Monitor.''}
\emph{European Centre for Medium-Range Weather Forecasts}.
\url{https://apps.climate.copernicus.eu/global-temperature-trend-monitor/}.

\bibitem[\citeproctext]{ref-Dadej2024}
Dadej, Mateusz. 2024. {``MarSwitching.jl: A Julia Package for Markov
Switching Dynamic Models.''} \emph{Journal of Open Source Software} 9
(98): 6441. \url{https://doi.org/10.21105/joss.06441}.

\bibitem[\citeproctext]{ref-CMIP6}
Eyring, V., S. Bony, G. A. Meehl, C. A. Senior, B. Stevens, R. J.
Stouffer, and K. E. Taylor. 2016. {``Overview of the Coupled Model
Intercomparison Project Phase 6 (CMIP6) Experimental Design and
Organization.''} \emph{Geoscientific Model Development} 9 (5): 1937--58.
\url{https://doi.org/10.5194/gmd-9-1937-2016}.

\bibitem[\citeproctext]{ref-GISTEMPTeam2020}
GISTEMP. 2020. {``{GISS Surface Temperature Analysis (GISTEMP), version
4}.''} \url{https://data.giss.nasa.gov/gistemp/}.

\bibitem[\citeproctext]{ref-albedo2023}
Goessling, Helge F., Thomas Rackow, and Thomas Jung. {``Recent Global
Temperature Surge Intensified by Record-Low Planetary Albedo.''}
\emph{Science} 0 (0): eadq7280.
\url{https://doi.org/10.1126/science.adq7280}.

\bibitem[\citeproctext]{ref-Granger1980Id}
Granger, C. W. J., and Roselyne Joyeux. 1980. {``{An Introduction to
Long‐Memory Time Series Models and Fractional Differencing}.''}
\emph{Journal of Time Series Analysis} 1 (1): 15--29.
\url{https://doi.org/10.1111/j.1467-9892.1980.tb00297.x}.

\bibitem[\citeproctext]{ref-Granger1980}
Granger, Clive W. J. 1980. {``Long Memory Relationships and the
Aggregation of Dynamic Models.''} \emph{Journal of Econometrics} 14:
227--38. \url{https://doi.org/10.1016/0304-4076(80)90092-5}.

\bibitem[\citeproctext]{ref-Haldrup2017}
Haldrup, Niels, and J. Eduardo Vera-Valdés. 2017. {``Long Memory,
Fractional Integration, and Cross-Sectional Aggregation.''}
\emph{Journal of Econometrics} 199 (1): 1--11.
\url{https://doi.org/10.1016/j.jeconom.2017.03.001}.

\bibitem[\citeproctext]{ref-nino2}
Ham, Yoo-Geun, Jeong-Hwan Kim, and Jing-Jia Luo. 2019. {``Deep Learning
for Multi-Year ENSO Forecasts.''} \emph{Nature} 573 (7775): 568--72.
\url{https://doi.org/10.1038/s41586-019-1559-7}.

\bibitem[\citeproctext]{ref-hamilton1989}
Hamilton, James D. 1989. {``A New Approach to the Economic Analysis of
Nonstationary Time Series and the Business Cycle.''} \emph{Econometrica:
Journal of the Econometric Society}, 357--84.

\bibitem[\citeproctext]{ref-hamilton1990}
---------. 1990. {``Analysis of Time Series Subject to Changes in
Regime.''} \emph{Journal of Econometrics} 45 (1-2): 39--70.

\bibitem[\citeproctext]{ref-nino4}
Hassanibesheli, Forough, Jürgen Kurths, and Niklas Boers. 2022.
{``Long-Term ENSO Prediction with Echo-State Networks.''}
\emph{Environmental Research: Climate} 1 (1): 011002.
\url{https://doi.org/10.1088/2752-5295/ac7f4c}.

\bibitem[\citeproctext]{ref-CM4}
Held, I. M., H. Guo, A. Adcroft, J. P. Dunne, L. W. Horowitz, J.
Krasting, E. Shevliakova, et al. 2019. {``Structure and Performance of
GFDL's CM4.0 Climate Model.''} \emph{Journal of Advances in Modeling
Earth Systems} 11 (11): 3691--3727.
\url{https://doi.org/10.1029/2019MS001829}.

\bibitem[\citeproctext]{ref-Hosking1981}
Hosking, J. R. M. 1981. {``Fractional Differencing.''} \emph{Biometrika}
68 (1): 165--76. \url{https://doi.org/10.1093/biomet/68.1.165}.

\bibitem[\citeproctext]{ref-Hurst1956}
Hurst, H. E. 1956. {``{The Problem of Long-Term Storage in
Reservoirs}.''} \emph{International Association of Scientific Hydrology.
Bulletin} 1 (3): 13--27.
\url{https://doi.org/10.1080/02626665609493644}.

\bibitem[\citeproctext]{ref-IPCC6WGI}
IPCC. 2021. \emph{{Summary For Policymakers. In: \emph{Climate Change
2021: The Physical Science Basis. Contribution of Working Group I to the
Sixth Assessment Report of the Intergovernmental Panel on Climate
Change.} {[}Masson-Delmotte, V., P. Zhai, A. Pirani, S.L. Connors, C.
Péan, S. Berger, N. Caud, Y. Chen, L. Goldfarb, M.I. Gomis, M. Huang, K.
Leitzell, E. Lonnoy, J.B.R. Matthews, T.K. Maycock, T. Waterfield, O.
Yelekçi, R. Yu, and B. Zhou (eds.){]}.}} Book. Cambridge University
Press, Cambridge, United Kingdom; New York, NY, USA.
\url{https://doi.org/10.1017/9781009157896.001}.

\bibitem[\citeproctext]{ref-Jiang2024}
Jiang, Ning, Congwen Zhu, Zeng-Zhen Hu, Michael J. McPhaden, Deliang
Chen, Boqi Liu, Shuangmei Ma, et al. 2024. {``Enhanced Risk of
Record-Breaking Regional Temperatures During the 2023--24 El Ni{ñ}o.''}
\emph{Scientific Reports} 14 (1): 2521.
\url{https://doi.org/10.1038/s41598-024-52846-2}.

\bibitem[\citeproctext]{ref-kunsch1987}
Künsch, Hans. 1987. {``Statistical Aspects of Self-Similar Processes.''}
\emph{Bernoulli} 1: 67--74.

\bibitem[\citeproctext]{ref-nino3}
L'Heureux, Michelle L., Aaron F. Z. Levine, Matthew Newman, Catherine
Ganter, Jing-Jia Luo, Michael K. Tippett, and Timothy N. Stockdale.
2020. {``ENSO Prediction.''} In \emph{El Niño Southern Oscillation in a
Changing Climate}, 227--46. American Geophysical Union (AGU).
\url{https://doi.org/10.1002/9781119548164.ch10}.

\bibitem[\citeproctext]{ref-MAGICC}
Meinshausen, M., S. C. B. Raper, and T. M. L. Wigley. 2011. {``Emulating
Coupled Atmosphere-Ocean and Carbon Cycle Models with a Simpler Model,
MAGICC6 -- Part 1: Model Description and Calibration.''}
\emph{Atmospheric Chemistry and Physics} 11 (4): 1417--56.
\url{https://doi.org/10.5194/acp-11-1417-2011}.

\bibitem[\citeproctext]{ref-HadCRUT5}
Morice, Colin P, John J Kennedy, Nick A Rayner, JP Winn, Emma Hogan, RE
Killick, RJH Dunn, TJ Osborn, PD Jones, and IR Simpson. 2021. {``An
Updated Assessment of Near-Surface Temperature Change from 1850: The
HadCRUT5 Data Set.''} \emph{Journal of Geophysical Research:
Atmospheres} 126 (3): e2019JD032361.

\bibitem[\citeproctext]{ref-MESMER}
Nath, S., Q. Lejeune, L. Beusch, S. I. Seneviratne, and C.-F.
Schleussner. 2022. {``MESMER-m: An Earth System Model Emulator for
Spatially Resolved Monthly Temperature.''} \emph{Earth System Dynamics}
13 (2): 851--77. \url{https://doi.org/10.5194/esd-13-851-2022}.

\bibitem[\citeproctext]{ref-GISS}
Orbe, Clara, David Rind, Jeffrey Jonas, Larissa Nazarenko, Greg
Faluvegi, Lee T. Murray, Drew T. Shindell, et al. 2020. {``GISS Model
E2.2: A Climate Model Optimized for the Middle Atmosphere---2.
Validation of Large-Scale Transport and Evaluation of Climate
Response.''} \emph{Journal of Geophysical Research: Atmospheres} 125
(24): e2020JD033151. \url{https://doi.org/10.1029/2020JD033151}.

\bibitem[\citeproctext]{ref-shipping2023}
Quaglia, I., and D. Visioni. 2024. {``Modeling 2020 Regulatory Changes
in International Shipping Emissions Helps Explain Anomalous 2023
Warming.''} \emph{Earth System Dynamics} 15 (6): 1527--41.
\url{https://doi.org/10.5194/esd-15-1527-2024}.

\bibitem[\citeproctext]{ref-jacob_quinn_2024_13955982}
Quinn, Jacob, Milan Bouchet-Valat, Nick Robinson, Bogumił Kamiński, Gem
Newman, Alexey Stukalov, Curtis Vogt, et al. 2024. {``JuliaData/CSV.jl:
V0.10.15.''} Zenodo. \url{https://doi.org/10.5281/zenodo.13955982}.

\bibitem[\citeproctext]{ref-Rohde2024}
Rohde, Robert. 2024. {``October 2024 Temperature Update.''}
\emph{Berkeley Earth}.
\url{https://berkeleyearth.org/october-2024-temperature-update/}.

\bibitem[\citeproctext]{ref-BerkeleyTemp}
Rohde, Robert A, and Zeke Hausfather. 2020. {``The Berkeley Earth
Land/Ocean Temperature Record.''} \emph{Earth System Science Data} 12
(4): 3469--79.

\bibitem[\citeproctext]{ref-Schneider2001}
Schneider, Stephen H. 2001. {``What Is 'Dangerous' Climate Change?''}
\emph{Nature} 411 (6833): 17--19.
\url{https://doi.org/10.1038/35075167}.

\bibitem[\citeproctext]{ref-Schneider2005}
Schneider, Stephen H., and Michael D. Mastrandrea. 2005.
{``Probabilistic Assessment of {`Dangerous'} Climate Change and
Emissions Pathways.''} \emph{Proceedings of the National Academy of
Sciences} 102 (44): 15728--35.
\url{https://doi.org/10.1073/pnas.0506356102}.

\bibitem[\citeproctext]{ref-Schneider2023}
Schneider, Tapio, Swadhin Behera, Giulio Boccaletti, Clara Deser, Kerry
Emanuel, Raffaele Ferrari, L. Ruby Leung, et al. 2023. {``Harnessing AI
and Computing to Advance Climate Modelling and Prediction.''}
\emph{Nature Climate Change} 13 (9): 887--89.
\url{https://doi.org/10.1038/s41558-023-01769-3}.

\bibitem[\citeproctext]{ref-BIC1978}
Schwarz, Gideon. 1978. {``Estimating the Dimension of a Model.''}
\emph{The Annals of Statistics} 6 (2): 461--64.
\url{http://www.jstor.org/stable/2958889}.

\bibitem[\citeproctext]{ref-shimotsu2005exact}
Shimotsu, Katsumi, and Peter CB Phillips. 2005. {``Exact Local Whittle
Estimation of Fractional Integration.''} \emph{The Annals of Statistics}
33 (4): 1890--1933.

\bibitem[\citeproctext]{ref-Thirumalai2017}
Thirumalai, Kaustubh, Pedro N. DiNezio, Yuko Okumura, and Clara Deser.
2017. {``Extreme Temperatures in Southeast Asia Caused by El Ni{ñ}o and
Worsened by Global Warming.''} \emph{Nature Communications} 8 (1):
15531. \url{https://doi.org/10.1038/ncomms15531}.

\bibitem[\citeproctext]{ref-nino1}
Thirumalai, Kaustubh, Pedro N. DiNezio, Judson W. Partin, Dunyu Liu,
Kassandra Costa, and Allison Jacobel. 2024. {``Future Increase in
Extreme El Ni{ñ}o Supported by Past Glacial Changes.''} \emph{Nature}
634 (8033): 374--80. \url{https://doi.org/10.1038/s41586-024-07984-y}.

\bibitem[\citeproctext]{ref-VERAVALDES2024a}
Vera-Valdés, J. E. 2024. {``LongMemory.jl: Generating, Estimating, and
Forecasting Long Memory Models in Julia.''} \emph{arXiv Preprint
arXiv:2401.14077}. \url{https://arxiv.org/abs/2401.14077}.

\bibitem[\citeproctext]{ref-Vera-Valdes2021b}
Vera-Valdés, J. Eduardo. 2021. {``{Temperature Anomalies, Long Memory,
and Aggregation}.''} \emph{Econometrics} 9 (1): 1--22.
\url{https://doi.org/10.3390/econometrics9010009}.

\bibitem[\citeproctext]{ref-Wigley2001}
Wigley, T. M. L., and S. C. B. Raper. 2001. {``Interpretation of High
Projections for Global-Mean Warming.''} \emph{Science} 293 (5529):
451--54. \url{https://doi.org/10.1126/science.1061604}.

\bibitem[\citeproctext]{ref-wooldridge2010econometric}
Wooldridge, Jeffrey M. 2010. \emph{Econometric Analysis of Cross Section
and Panel Data}. MIT press.

\bibitem[\citeproctext]{ref-Zaffaroni2004}
Zaffaroni, Paolo. 2004. {``Contemporaneous Aggregation of Linear Dynamic
Models in Large Economies.''} \emph{Journal of Econometrics} 120:
75--102. \url{https://doi.org/10.1016/S0304-4076(03)00207-0}.

\end{CSLReferences}
\end{document}